\documentstyle[12pt]{article}

\newcommand{\be}{\begin{equation}}
\newcommand{\ee}{\end{equation}}
\newcommand{\ba}{\begin{eqnarray}}
\newcommand{\ea}{\end{eqnarray}}

\begin{document}
\def\pct#1{(see Fig. #1.)}

\begin{titlepage}
\hbox{\hskip 12cm CERN-TH/96-99  \hfil}
\hbox{\hskip 12cm ROM2F-96/20  \hfil}
\hbox{\hskip 12cm hep-th/9604097 \hfil}
\vskip 1.4cm
\begin{center} 
{\Large  \bf  Low-Energy \ Analysis \ of \ $M$ \ and \ $F$ \ 
Theories \vskip .6cm
on \ Calabi-Yau \ Threefolds}
 
\vspace{1.8cm}
 
{\large \large Sergio Ferrara$^{\dagger}$, \ \ Ruben Minasian$^{\dagger}$ 
\ \ and \ \ Augusto Sagnotti$^{*}$}

\vspace{0.6cm}

{$^{\dagger}$ \sl Theory \ Division, \ \ CERN  \\ CH-1211  Geneva 23, \ \ 
SWITZERLAND}

\vspace{0.3cm}

{$^{*}$ \sl Dipartimento di Fisica, \ \
Universit{\`a} di Roma \ ``Tor Vergata'' \\
I.N.F.N.\ - \ Sezione di Roma \ ``Tor Vergata'', \ \
Via della Ricerca Scientifica , 1 \\
00133 \ Roma, \ \ ITALY}
\end{center}
\vskip 1.5cm

\abstract{
We elucidate the interplay between gauge and supersymmetry anomalies
in six-dimensional $N=1$ supergravity with generalized couplings between  
tensor and vector multiplets.  We derive the structure of the five-dimensional
supergravity resulting from the $S_1$ reduction of these models and give the
constraints on Chern-Simons couplings that follow from duality to $M$ theory 
compactified on a Calabi-Yau threefold.  The duality is supported only on a 
restricted class of Calabi-Yau threefolds and requires a special type of 
intersection form. We derive  five-dimensional central-charge formulas and 
briefly discuss the associated phase transitions. Finally, we exhibit 
connections with $F$-theory compactifications on Calabi-Yau 
manifolds that admit elliptic fibrations. This analysis suggests that $F$ 
theory unifies Type-$IIb$ three-branes and $M$-theory five-branes.} 
\vfill
\hbox{\hskip 1.2cm CERN-TH/96-99  \hfil}
\hbox{\hskip 1.2cm April 1996 \hfil}
\end{titlepage}
\makeatletter
\@addtoreset{equation}{section}
\makeatother
\renewcommand{\theequation}{\thesection.\arabic{equation}}
\addtolength{\baselineskip}{0.3\baselineskip} 
\section{Introduction}

Six-dimensional $N=1$ supergravity theories  with an arbitrary 
number of tensor multiplets \cite{rom, as} 
arise very naturally in open descendants \cite{cargese} 
of Type-$IIb$ $K_3$ reductions \cite{bs}.
They also correspond to non-perturbative heterotic vacua \cite{dmw,sw},
and play a role in orbifold compactifications of $M$ theory \cite{sen,morb} and in
Calabi-Yau (CY) compactifications of $F$ theory \cite{vafaf,morvafa}.  In this paper we
discuss the low-energy field theory of these models both from a five- and 
from a six-dimensional
viewpoint, relating them to eleven-dimensional supergravity \cite{cjs} (the low-energy limit of
$M$ theory) compactified on CY  manifolds, as well as to $F$ theory.

Upon circle compactification, six-dimensional $N=1$ supergravity with $n_T$ tensor multiplets,
$n_V$ vector multiplets and $n_H$ hypermultiplets yields five-dimensional simple
supergravity with $n_T + n_V + 1$ vector multiplets and $n_H$ hypermultiplets. The resulting
five-dimensional interactions of the vector multiplets are entirely specified by the
Chern-Simons couplings of the $n_T + n_V + 2$ vector fields \cite{gst},
\be
C_{\Lambda \Sigma \Delta} \ \int \ A^{\Lambda} \wedge F^{\Sigma} \wedge F^{\Delta} \quad .
\label{fivecs}
\ee
In compactifications of eleven-dimensional supergravity on CY threefolds, the 
symmetric constants $C_{\Lambda \Sigma \Delta}$ are the intersection 
numbers. The
vector fields arise from the expansion of the three-form tensor in the $H^{1,1}$
cohomology, whose dimension is related to the number of massless multiplets
by $h^{1,1} = n_V + n_T + 2$. The CY volume deformation belongs to a five-dimensional
universal hypermultiplet, so that
$n_H = h^{2,1} + 1$.  This is also the counting for six-dimensional vacua obtained from $F$
theory on the same CY manifold.  Moreover, the vanishing condition for the irreducible part
of the six-dimensional gravitational anomaly, $n_H \ - \ n_V \ + \ 29 \ n_T \ = \ 273$, requires
that
\be
n_T = 9 + {1 \over 60} \ \chi \quad ,
\label{euler}
\ee
where $\chi = 2 ( h^{1,1} - h^{2,1} )$ is the Euler characteristic of the CY threefold.
This relation holds when all six-dimensional vectors are Abelian.

When the five-dimensional theory is obtained by $S_1$ compactification from six dimensions,
the intersection numbers
 $C_{\Lambda \Sigma \Delta}$ are subject to certain restrictions,
which should be satisfied in $F$-theory constructions and are necessary 
constraints on CY
threefolds in order that a six-dimensional interpretation be possible. In particular, CY
threefolds that are elliptic fibrations \cite{morvafa} should fall in this class. The
restrictions generalize those that associate $M$-theory compactifications with heterotic duals
to CY threefolds that are $K_3$ fibrations.  In these cases, the effective field
theory contains a preferred vector that may be turned into an antisymmetric tensor
of the dual heterotic theory.  In the more general case of $n_T > 1$, we shall see that the
restrictions on the CY threefold allow a total of $n_T$ such preferred vectors.
The restrictions overlap with those resulting from $K_3$ fibrations, but in general they
differ.  In particular, when $n_V = 0$ the five-dimensional vector multiplet moduli space is
$O(1,1) \times {{O(1,n_T)} \over {O(n_T)}}$, a result inherited from the six-dimensional
couplings of  \cite{rom}.  This property will play a crucial role in the $F$-theory
interpretation of the $(11,11)$ CY threefold, the simplest example exhibiting these features.
When $n_V > 0$ and $n_T \ge 1$, the moduli space is no longer homogeneous (aside from the
special case $n_T = 1$, and in the absence of certain Chern-Simons couplings for vector
multiplets, where one can obtain $O(1,1) \times {{O(1,n_V + 1)} \over {O(n_V + 1)}}$), and the
theory will generally undergo phase transitions for finite values of the moduli \cite{as,dmw}, 
as already occurs in six dimensions.

The plan of this paper is as follows.
In Section 2 we extend the results of  \cite{as} on the supersymmetric coupling of
tensor multiplets and YM multiplets.  In particular, we elucidate the interplay between
gauge and supersymmetry anomalies in these theories and show that, 
already at lowest order in the Fermi fields, the
anomalous supersymmetry Ward identities lead to Yang-Mills currents different from
those of  \cite{as}, which embody the consistent form of the gauge anomaly. 
For the case of a single tensor multiplet coupled to supergravity, or more generally
for (non-supersymmetric) models with several tensors not restricted by (anti)self-duality,
the resulting equations follow from a Lagrangian, while the field 
equations for
several tensor multiplets coupled to supergravity are nicely determined by the $O(1,n_T )$
symmetry.  Despite the presence of anomalies,
supersymmetry retains its predictive power, since the Wess-Zumino consistency conditions 
\cite{wz} link gauge and supersymmetry anomalies.  A corresponding phenomenon 
occurs in globally supersymmetric four-dimensional models 
\cite{wzsusy}.  In Section 3 we consider the reduction to five dimensions. This results in the
standard form of five-dimensional supergravity \cite{gst}, since the left-over anomaly,
cohomologically trivial, may be disposed of by a local counterterm involving solely the $n_V$
vector multiplets.  In Section 4 we derive central-charge formulas for five-dimensional
electric (point-like) and magnetic (string-like) states.  The six-dimensional phase transitions
are accompanied by additional ones that are briefly discussed.  In Section 5 we compare
our results to $F$-theory  constructions and analyze
the effective Lagrangians of some CY threefolds.  Certain models (e.g.
the (11,11) CY threefold) are suggestive of dual descriptions related by 
the interchange of $( n_T - 1 )$ and $n_V$.  Finally, in Section 6 we display
some twelve-dimensional geometrical couplings\footnote{Geometrical couplings do not involve
the space-time metric. Topological couplings are a subset of these that give vanishing
results upon integration over topologically trivial manifolds.} that suggest that $F$ theory
unifies the Type-$IIb$ three-brane with the $M$-theory five-brane.

\vskip 24pt
\section{Tensor and Vector Multiplets in Six-Dimensional Supergravity}

In this section we extend the construction of  \cite{as}, elucidating some
features of the resulting field equations that may play a role in future constructions.  In
particular, we display a new phenomenon in supergravity, whereby gauge anomalies reflect
themselves in suitable supersymmetry anomalies, fully determined by the Wess-Zumino consistency
conditions \cite{wz}.  The analogue of this phenomenon in globally supersymmetric models was
discussed in  \cite{wzsusy}.

A generalized Green-Schwarz mechanism \cite{gs} in
six-dimensional supergravity, involving several tensor and vector multiplets,
was motivated by the systematic appearance, first noted in 
\cite{bs}, of several tensor multiplets in open descendants \cite{cargese}
of Type-$IIb$ $K_3$ compactifications. The coupling of several tensor multiplets to simple
six-dimensional supergravity
had been studied previously by Romans \cite{rom}, as an interesting 
generalization
of the methods of  \cite{type2b} to models with scalar fields,
while the coupling of vector multiplets to six-dimensional supergravity with
a single tensor multiplet was originally considered in  \cite{ns}.  Couplings
related to  \cite{as} have recently been displayed in  \cite{sw} for models 
with a single
tensor multiplet, where a Lagrangian can be explicitly constructed.

After all tadpole constraints
were enforced in the open descendants of \cite{bs},
the residual anomaly polynomial revealed, upon diagonalization, an
$O(1,n_T)$-like structure of the type
\be
A \ \sim \ \sum_{r,s} \ \eta_{rs} \ c^{r z} \ c^{s {z^{\prime}}} \ Tr_z (F^2) \ 
Tr_{z^{\prime}} (F^2) \quad . \label{anom}
\ee
As in  \cite{as}, $\eta_{rs}$ denotes a Minkowski metric of signature 
$(1-n_T)$, and $z$ labels
the various simple factors of the gauge group, while the $c_z^r$ are rescaled in all
component expressions.  In all
the models of
\cite{bs} that have been analyzed, the gravitational contribution is 
confined to the $r=0$ term. 
It is of higher order in the derivatives, and thus we shall ignore it as in 
\cite{as}, although
we shall return to it at the end of this section.

Whereas the conventional Green-Schwarz mechanism \cite{gs} does not apply directly to these
models, the residual anomaly may be removed by the combined action of several
antisymmetric tensors, as dictated by the constants $c^r_z$.  In diagonal rational
open-string models these have a microscopic interpretation in terms of
the $S$ matrix of the conformal theory \cite{as}, while in more general rational models they
are related to the tensors $A$ introduced in  \cite{pss}.
One is thus led to consider generalized Chern-Simons couplings of antisymmetric
tensors valued in the vector representation of $O(1,n_T)$, with field strengths
\be
H^r \ = \ d B^r \ -  \ c^{r z} \ \omega_z \quad ,
\label{hr}
\ee
where $\omega_z$ are Chern-Simons three-forms for the gauge
fields of the vector multiplets.  As usual, the gauge invariance of $H^r$ \cite{tendsugra}
demands that
$B^r$ change under vector gauge transformations according to
\be
\delta B^r \ =  \ c^{r z} \ Tr_z ( \Lambda dA ) \quad .
\label{deltab}
\ee

It should be appreciated that in these models the Chern-Simons couplings are 
induced by the residual gauge anomaly. This feature will be
reflected in the resulting field equations, that embody the residual 
gauge anomaly.
As we shall see, the latter is quadratic in the $c^{r z}$, that draw their origin
naturally from genus-$1 \over 2$ open-string amplitudes \cite{as}.

Together with the antisymmetric tensor fields $B^r$ and the gauge fields 
$A$, the low-energy
field theory includes the vielbein and the scalar field coordinates 
$v_r$ of ${{O(1,n_T )} \over {O(n_T )}}$, that satisfy the quadratic 
constraint \be
{\cal V} \ = \ \eta_{rs} \ v^r \ v^s \ = \ v^r \ v_r \ = \ 1 \quad .
\label{scalars}
\ee
The additional elements of the scalar $O(1,n_T)$ matrix, denoted by $x_r^m$ in
\cite{rom,as}, satisfy the constraints
\be
v_r \ v_s \ - \ x_r^m \ x_s^m \ = \ \eta_{rs} \qquad \qquad \eta^{rs} \  x_r^m \ x_s^n \ = \ -
\ \delta^{mn}
\label{constraints}
\ee
and enter the composite $O(n_T)$ connection.
The fermionic fields are a pair of left-handed gravitini $\psi_{\mu}$, $n_T$ 
pairs of right-handed spinors $\chi^m$ and pairs of left-handed gaugini $\lambda$, all
satisfying symplectic Majorana-Weyl conditions.  As in  \cite{as}, we work with
a space-time metric of signature $(1-5)$, restricting our attention to terms of
lowest order in the Fermi fields.

In order to simplify
the equations for the bosonic fields, naturally neutral with respect to the local
composite  $O(n_T)$ symmetry, it is convenient to introduce the matrix
\be
G_{rs} = \ - \ {1 \over 2} \ \partial_r \partial_s \ {\rm log} ( v^r \ v_r ) \ 
{\big \vert}_{\cal V} \ = \
2 \ v_r \ v_s \ - \ \eta_{rs} \quad , \label{G}
\ee
since it reduces the (anti)self-duality constraints for the antisymmetric tensors to 
\be
G_{r s} \ {H^s}^{\mu \nu \rho} \ = \  \eta_{rs} \ *{H^s}^{\mu \nu \rho} \ = \  
{1 \over {6 e}} \ \epsilon^{\mu \nu \rho \alpha \beta \gamma} \ 
{H_r}_{\alpha \beta \gamma} \quad .
\label{sdual}
\ee
The second-order tensor equation of  \cite{as},
\be
\nabla_{\mu} \left( G_{rs} H^{s \mu \nu \rho} \right) \ = \
- \ {1 \over {4 e}} \ \epsilon^{\nu \rho \alpha \beta \gamma \delta} \ c_r^z \ Tr_z 
\left( F_{\alpha \beta} \ F_{\gamma \delta} \right) \quad ,
\label{divH}
\ee
also takes a simpler form, since now its source does not involve the scalar fields.
This equation follows from the Bianchi identity 
for $H^r$ resulting from eq. (\ref{hr}) and from the (anti)self-duality
condition of eq. (\ref{sdual}). Tensor current conservation is implied by the Bianchi
identity for the vector field strengths.

To lowest order, the fermionic field equations  are
\ba
{\gamma}^{\mu \nu \rho} \ D_{\nu} \ \psi_{\rho}  &+& 
v_r \ H^{r \mu \nu \rho} \ \gamma_{\nu} \ \psi_{\rho} \ -  \ { i \over 2} 
\ x_r^m \
H^{r \mu \nu \rho} \ \gamma_{\nu \rho} \ \chi^m \  \label{rs} \\ \nonumber 
&+& {i \over 2} \ x_r^m \ \partial_{\nu} v^r \ \gamma^{\nu} \ \gamma^{\mu} \
 \chi^m \
- \  {1 \over {\sqrt{2}}} \ \gamma^{\sigma \tau} \ {\gamma^{\mu}} \
v_r \ c^{rz} \ {\rm tr}_z ( F_{\sigma \tau} \ \lambda ) \ = \ 0 
\ea
\ba
\gamma^{\mu} \ D_{\mu} \ \chi^m &-& {1 \over 12} \ v_r \ {H^r}_{\mu \nu \rho} \
\gamma^{\mu \nu \rho} \ \chi^m \ - \ {i \over 2} \ x_r^m \ H^{r \mu \nu \rho} \
\gamma_{\mu \nu} \ \psi_{\rho} \label{chim} \\ \nonumber
&-&  {i \over 2} \
x_r^m \ \partial_{\nu} v^r \ {\gamma}^{\mu} \ {\gamma}^{\nu} \ \psi_{\mu} \
- \ {i \over {\sqrt{2}}} \ {x^m}_r \ c^{rz} \ {\rm tr}_z ( \gamma^{\mu \nu}
\lambda
\ F_{\mu \nu} )
\ = \ 0  
\ea
\ba
v_r \ c^{rz} \ \gamma^{\mu} \ D_{\mu} \lambda  &+&
{1 \over 2} \ c^{rz} \ \partial_{\mu} v_r \ \gamma^{\mu} \ \lambda  
\ + \
{1 \over {2 \sqrt{2}}} \ ( v_r \ c^{rz}) \ F_{\lambda \tau} \ \gamma^{\mu} \
\gamma^{\lambda \tau} \ \psi_{\mu} \label{gaugini} \\ \nonumber 
&+& \ {i \over {2 \sqrt{2}}} \ ({x^m}_r \ c^{rz})
\
\gamma^{\mu \nu} \ \chi^m \ F_{\mu \nu} \ - \ {1 \over {12}} \ c^{rz} \ {H^r}_{\mu \nu \rho} 
\ \gamma^{\mu \nu \rho} \ \lambda \ = \ 0  \quad , 
\ea
where the last term in eq. (\ref{gaugini}) is (inexplicably) absent in  \cite{as}. 

All bosonic couplings in the bosonic field equations may be obtained from 
the supersymmetry
variations of eqs. (\ref{rs}), (\ref{chim}) and (\ref{gaugini}), making 
use of the
first- and second-order tensor equations. Thus, the scalar field equation is
\be
x_r^m \ \nabla_{\mu} \ \partial^{\mu} \ v^r \ + \ {2 \over 3} \ x_r^m \ v_s \ H^{r \mu \nu \rho} 
{H^s}_{\mu \nu \rho} \ - \  \ x_r^m \ c^{r z} \ Tr_z ( F_{\alpha \beta} \
F^{\alpha \beta} )  \ = \ 0 \quad , \label{scalar}
\ee
where the overall $x_r^m$ reflects the constraint of eq. (\ref{scalars}),
while Einstein's equations are
\ba
 R_{\mu \nu} \ &-& \ {1 \over 2} \ g_{\mu \nu} \ R \ + \  \partial_{\mu} v^r \ \partial_{\nu} v_r
- {1 \over 2} \ g_{\mu \nu} \ \partial^{\rho} v^r \ \partial_{\rho} v_r  \nonumber \\
&-&  G_{rs} \ {H^r}_{\mu \rho \sigma} \
{{H^s}_{\nu}}^{\rho \sigma} \ + \
4 \ v_r \ c^{rz} \ {\rm tr}_z ( F_{\lambda \mu} \ {F^{\lambda}}_{\nu} \ - \
{1 \over 4} g_{\mu \nu} \ F^2 ) \ = \ 0  \quad .
\label{einstein}
\ea
Finally, the vector field equation is
\be
D_{\mu} ( v_r \ c^{rz} \  F^{\mu \nu} ) \ - \ c^{rz} \ G_{rs} \ H^{s \nu \rho \sigma} \
F_{\rho
\sigma} \ = \ 0 \quad .
\label{vector}
\ee
Realizing the supersymmetry algebra
on the fields requires a modification of the tensor transformation \cite{as}, and the
resulting supersymmetry transformations are
\ba
&& \delta_{\epsilon} \ {e_{\mu}}^m \ = \ - \ i \ \bar{\epsilon} \ \gamma^m
\ \psi_{\mu} \nonumber \\
&& \delta_{\epsilon} \ \psi_{\mu} \ = \ D_{\mu} \ \epsilon \ + \ {1 \over 4}
\ v_r \ {H^r}_{\mu \nu \rho} \ \gamma^{\nu \rho}  \epsilon \nonumber \\
&& \delta_{\epsilon} \ {B^r}_{\mu \nu} \ = \ i \ v^r \ {\bar{\psi}}_{[ \mu}
\gamma_{ \nu ]}  \epsilon \ + \ {1 \over 2} \ x^{rm} \
{\bar{\chi}}^m  \ \gamma_{\mu \nu} \epsilon 
\ - \ 2 \ c^{rz} \ {\rm {tr}_z} ( A_{[ \mu} \ \delta_{\epsilon}
        A_{\nu ]} ) \nonumber \\
&& \delta_{\epsilon} \ \chi^m \ = \ + \ {i \over 2} \ \partial_{\mu} v^r 
\ x_r^m \ {\gamma}^{\mu} \
\epsilon \ + \ {i \over 12} \ x_r^m \ {H^r}_{\mu \nu \rho} \ 
{\gamma}^{\mu \nu \rho} \epsilon \nonumber \\ 
&& \delta_{\epsilon} \ v_r \ = \ {x^m}_r \ \bar{\epsilon} \ {\chi}^m \nonumber \\
&& \delta_{\epsilon} \ \lambda \ = \ - \ {1 \over {2 \sqrt{2}}} \ F_{\mu \nu} \
{\gamma}^{\mu \nu} \epsilon \nonumber \\
&& \delta_{\epsilon} \ A_{\mu} \ = \ - \ {i \over {\sqrt{2}}} ( \bar{\epsilon} \ \gamma_{\mu}
\ \lambda )
\label{susy}
\ea

Both the field equations and the supersymmetry transformations should be completed
by the addition of higher-order spinor terms.  At any rate, it may be 
verified that, on all
bosonic fields, the commutator of two supersymmetry transformations closes on
{\it all} local symmetries
\ba
[ \delta_{\epsilon_1} , \delta_{\epsilon_2} ]  & = & \delta_{\rm gct} ( \xi^{\mu} = -
i \bar{\epsilon}_1 \gamma^{\mu} \epsilon_2 ) + \delta_{\rm tens}( \Lambda_{\mu}^r =
- {1 \over 2} v^r \xi^{\mu} - \xi^{\nu} B_{\mu \nu}^r ) 
+ \delta_{\rm vect}( \Lambda = - \xi^{\mu} A_{\mu} ) \nonumber \\ 
& + & \delta_{\rm susy}( \zeta =  - i \xi^{\mu} \psi_{\mu} ) +  
\delta_{\rm Lorentz}( \Omega^{mn} = \xi^{\mu} ( {\omega}_{\mu}^{mn} -
v_r {H}^{rmn}_{\mu} ) ) \quad .
\label{susyalg}
\ea
Again, in deriving this result one must use the first-order equation for the tensor fields.  In
addition, the torsion equation contains a contribution from the gaugini:
\ba
D_{\mu} \ e_{\nu}^m \ - \ D_{\nu} \ e_{\mu}^m \ &+& \ i \ ( \bar{\psi}_{\mu} \
\gamma^{m} \ \psi_{\nu} ) \ - \ {i \over 4} \ ( {\bar{\chi}^m} \
{\gamma_{\mu \nu}}^{m} \ {\chi^m} ) \nonumber \\
&+& \ {i \over 2} \ v_r \ c^{rz} \ Tr_z( \bar{\lambda}
\ \gamma_{\mu \nu \rho} \ \lambda ) \ = \ 0 \qquad ,
\label{torsion}
\ea
while some supercovariantizations are implicit.

Equation (\ref{vector}) is quite peculiar.  First of all, as noted in  
\cite{as}, it could be derived from the action
\be
{\cal L}_v \ = \ - \ {e \over 2} \ v_r \ c^{rz} \ Tr_z ( F^{\mu \nu} F_{\mu \nu} ) \quad ,
\label{vectokin}
\ee
and thus the kinetic terms are positive {\it only} if the 
scalar fields are restricted to particular subregions of the moduli space, delimited by
boundaries where the effective gauge couplings diverge.  This,
however, is a blessing in disguise, since as proposed in  \cite{dmw},  
the singularity signals a phase transition, which may, as in 
\cite{sw,dp}, be ascribed to tensionless strings, and helps one gain 
insight into the structure of six-dimensional vacua.  
The second unusual feature of eq. (\ref{vector}) is that the vector gauge currents 
\be
J^{\mu} \ = \ 2 \ c^{rz} \ G_{rs} \ H^{s \mu \rho \sigma} \ F_{\rho \sigma}
\label{jcov}
\ee
that result from the coupling to the tensor multiplets are in general {\it not} conserved, as
pertains to a theory with an anomaly to be disposed of by fermion loops.
This is the reducible part of the six-dimensional gauge anomaly, the portion left-over after
tadpole conditions are enforced in open-string loop amplitudes as in  \cite{bs}.
Indeed, taking the divergence of eq. (\ref{jcov}) and making use of eq. (\ref{divH})
one finds
\be
D_{\mu} \ J^{\mu} \ = \ - \ {1 \over {2 e}} \ \epsilon^{\mu \nu \alpha \beta
\gamma \delta} \ c^{rz} \ c_r^{z^{\prime}} \ F_{\mu \nu} \ Tr_{z^{\prime}} ( F_{\alpha \beta}
F_{\gamma \delta} ) \quad ,
\label{covan}
\ee
that may be recognized as the covariant form of the residual anomaly.  It involves the
$O(1,n_T)$ lorentzian product of pairs of $c^r_z$ coefficients, a natural measure of the
chirality of the tensor couplings.

It is interesting to ask whether one can arrive at a vector equation
embodying the consistent form of the residual anomaly.  That this should be possible is
suggested by the long-held expectation that covariant and consistent anomalies are related by
local counterterms \cite{bz}.  Still, the latter form is more satisfactory, and pursuing this
issue is quite instructive, since the solution of the problem rests on
peculiar properties of the supersymmetry algebra. These have already emerged
in globally supersymmetric models \cite{wzsusy}.  The basic observation is that these field
equations embody a vector gauge anomaly and, as is usually the case in component
formulations, the commutator of two supersymmetry transformations in eq. (\ref{susyalg})
involves the anomalous vector gauge transformations. 

Even without a Lagrangian formulation, by combining the field equations with 
the corresponding
supersymmetry variations one can retrieve the total variation of the 
effective action under
local supersymmetry. Denoting by ${\cal A}_{\Lambda}$ the vector gauge anomaly and by
${\cal A}_{\epsilon}$ the supersymmetry anomaly, one has the Wess-Zumino 
consistency conditions:
\ba
& &\delta_{\epsilon_1} \ {\cal A}_{\epsilon_2} \ - \ \delta_{\epsilon_2} \ {\cal A}_{\epsilon_1}
\ = \ {\cal A}_{\Lambda(\epsilon_1 , \epsilon_2 )} \quad , \nonumber \\
& &\delta_{\Lambda} \ {\cal A}_{\epsilon} \ = \ \delta_{\epsilon} \ {\cal A}_{\Lambda} \quad ,
\label{wz}
\ea
and the first clearly requires a non-vanishing supersymmetry anomaly.  
Here we confine
our attention to the bosonic contributions to the vector current, although 
a systematic use of eqs.
(\ref{wz}) would also determine the fermionic terms in ${\cal A}_{\Lambda}$ and ${\cal
A}_{\epsilon}$, some of which may be anticipated from  \cite{wzsusy}. 

With the expected residual bosonic contribution to the vector gauge anomaly
\be
{\cal A}_{\Lambda} \ = \ \gamma \ \epsilon^{\mu \nu \alpha \beta
\gamma \delta} \ c^{rz} \ c_r^{z^{\prime}} \ Tr_z ( \Lambda \ \partial_{\mu} A_{\nu} ) \
Tr_{z^{\prime}} ( F_{\alpha
\beta} F_{\gamma \delta} ) \quad ,
\label{agauge}
\ee
one may verify that the second of eqs. (\ref{wz}) determines the relevant part of the 
supersymmetry anomaly,
\ba
{\cal A}_{\epsilon} \ = \ &-& \gamma \ \epsilon^{\mu \nu \alpha \beta
\gamma \delta} \ c^{rz} \ c_r^{z^{\prime}} \ Tr_z ( \ A_{\mu} \delta_{\epsilon} A_{\nu} )
\ Tr_{z^{\prime}} ( F_{\alpha
\beta} F_{\gamma \delta} ) \nonumber \\
&-& \ 4 \ \gamma \ \epsilon^{\mu \nu \alpha \beta
\gamma \delta} \ c^{rz} \ c_r^{z^{\prime}} \ Tr_z (\delta_{\epsilon} A_{\nu} F_{\mu \alpha} )
\ {\omega}_{z^{\prime} \beta \gamma \delta}  \quad .
\label{aususy}
\ea
Whereas the other Wess-Zumino consistency condition of eq. (\ref{wz}) would
fix this overall factor as well, it is simpler to fix it from the divergence
of the gauge current. Indeed, while combining eq. (\ref{vector}) with the other
field variations would result
in an effective action invariant under supersymmetry, demanding that the total variation be
${\cal A}_{\epsilon}$ alters the gauge current, and in particular the choice $\gamma =
-1/4$ reproduces precisely the consistent anomaly of eq. (\ref{agauge}):
\be
D_{\mu} \ \hat{J}^{\mu} \ = \ - \ {1 \over {4 e}} \ \epsilon^{\mu \nu \alpha \beta
\gamma \delta} \ c^{rz} \ c_r^{z^{\prime}} \ \partial_{\mu} A_{\nu} \ Tr_{z^{\prime}} (
F_{\alpha \beta} F_{\gamma \delta} ) \quad .
\ee

This hints to a simpler way of obtaining these
results, as well as to the rather general nature of the phenomenon, with potential applications
in other contexts. As in the ten-dimensional Green-Schwarz mechanism
\cite{gs}, one is embodying in the dynamics a portion of the gauge anomaly that the antisymmetric
tensors make cohomologically trivial.  Differently from that case, however, the resulting
modifications are already present at the lowest order in the derivative 
expansion. The
construction of field theories with these anomalies thus acquires some interest
of its own, and one may dispense momentarily with the restriction to (anti)self-dual tensors
introduced by six-dimensional supersymmetry to consider Lagrangians of 
the type \ba
{\cal L} \ &=& \ {e \over {12}} \ G_{rs} \ H^{r \mu \nu \rho} \ {H^s}_{\mu \nu \rho} - {e \over 2}
\ v_r \ c^{rz} \ Tr_z ( F_{\mu \nu} F^{\mu \nu} )  \label{lagrangian} \\
&-& {1 \over {8}} \ \epsilon^{\mu \nu \alpha \beta
\gamma \delta} \ {\tilde{c}}_r^z \ B^r_{\mu \nu} \ Tr_{z} (
F_{\alpha \beta} F_{\gamma \delta} ) 
\ - \ {1 \over {2}} \ \epsilon^{\mu \nu \alpha \beta
\gamma \delta} \ c_r^z \ {\tilde{c}}^{r z^{\prime}} \ \omega_{z \mu \nu \alpha} \
\omega_{z^{\prime} \beta \gamma \delta}  \nonumber
\ea
for unconstrained tensor fields, with Chern-Simons couplings as in eq. (\ref{hr}) and
geometrical couplings
with coefficients $\tilde{c}^r$ in general different from the $c^r$ that
enter the tensor Bianchi identities.  The relative normalization of the last two
contributions is fixed by symmetry, and the resulting vector gauge current has the 
consistent anomaly
\be
{{\cal A}^{\prime}}_{\Lambda} \ = \ - \ {1 \over 8} \ \epsilon^{\mu \nu \alpha \beta
\gamma \delta} \ ( c^{rz} \ {\tilde{c}}_s^{z^{\prime}} + {\tilde{c}}^{rz} \ c_s^{z^{\prime}} ) \
Tr_z ( \Lambda \ \partial_{\mu} A_{\nu} ) \ Tr_{z^{\prime}} ( F_{\alpha
\beta} F_{\gamma \delta} ) \quad .
\label{acctilde}
\ee
The last term in eq. (\ref{lagrangian})  was considered in  \cite{dm}.
Although clearly vanishing for a single gauge factor or for $c_z^r$ 
proportional to
${\tilde{c}}_z^r$, it restores the symmetry of the anomaly.
For a single tensor multiplet coupled to supergravity, the two different terms in eq.
(\ref{lagrangian}) reflect the existence of two second-order invariant tensors in $O(1,1)$,
$\eta_{rs}$ and
$\epsilon_{rs}$, a property not shared by larger $O(1,n_T)$ groups, where the restriction to
(anti)self-dual tensors would imply the equality of the two sets of couplings 
and the vanishing of the additional contribution.

The inclusion of gravitational Chern-Simons
couplings would result in the presence, in eq. (\ref{lagrangian}), of additional terms
involving the gravitational curvature. These higher-derivative terms 
could be accommodated
extending the range of $z$ and treating the gravitational sector as an additional factor
of the gauge group, with corresponding $c^r$ coefficients clearly displayed in the
residual anomaly polynomial.  In all the models of \cite{bs} that have been
analyzed explicitly, the gravitational contribution is confined to the 
$r=0$ term of the
anomaly polynomial and couples to the sum of the vector contributions.

\vskip 24pt
\section{Reduction to Five Dimensions and $M$ Theory on a CY Threefold}

The five-dimensional action resulting from a CY compactification of $M$ theory includes
geometrical interactions between the $h^{1,1}$ vectors arising from the 
eleven-dimensional three-form \cite{cadcdf,antft,ferkm,wittfm}
\be
I_5^{geom}=- {1 \over 12} \ C_{\Lambda\Sigma\Delta}\int_{{\cal M}_5}
A_1^\Lambda \wedge F_2^\Sigma \wedge F_2^\Delta \quad .
\label{topcop}
\ee
The intersection numbers $C_{\Lambda\Sigma\Delta}$ determine the metric 
on the vector moduli space \cite{gst}.

The requirement that this theory  be dual to the reduction of a 
six-dimensional theory with an arbitrary number of tensor and vector 
multiplets  restricts the form of this interaction, and thus the 
intersection form of the CY manifold. Special properties of the 
intersection form simplify the analysis significantly. First, it is a
polynomial, and the different contributions are additive. Different special 
regimes can be analyzed separately and then joined
together in the general expression. Second, it is a set of numbers, 
moduli-independent, 
and thus all scalar fields can be set to constant values when analyzing it.

Our strategy will be to analyze in some detail the reduction of the  $n_T 
=1$  Lagrangian and 
then extend the results in an $O(1, n_T)$ covariant fashion to the $n_T > 
1$ case. In order to simplify the comparison 
with $M$ theory, we perform the reduction at a generic point of the
moduli space where all vector fields are Abelian.
The coefficients $c_z^r$ now become $c_{\rm xy}^r$, where ${\rm x,y} 
=1,...,n_V.$  For $n_T =1$, $v_{\rm xy}$ and ${\tilde v_{\rm xy}}$ are 
$c_{0 {\rm xy}} \pm c_{1 {\rm xy}}$ respectively and, following  \cite{sw}, we define
the $O(1,1)$ ``light-cone'' decomposition
\ba
\omega \ &=& \ \sum \ v^i \ \omega_i \quad , \nonumber \\
\tilde{\omega} \ &=& \ \sum \ \tilde{v}^i \ \omega_i \quad .
\label{omegas}
\ea
The standard compactification ansatz for the vielbein,
\be
e^{\hat a}_{\hat \mu} = \left(\matrix{e^a_{\mu}& r Z_{\mu} \cr
0& r \cr} \right)  \quad ,
\label{vierb}
\ee
determines the part of the five-dimensional Lagrangian
originating from the 
six-dimensional tensor multiplets and their interactions,
\ba
{\cal L}_{5} \ &=& \ {r \over 2} \ e^{-2 \phi} \ \bigl( H \ - \ H^6 \wedge Z \bigr)
\wedge * \bigl( H \ - \ H^6 \wedge Z  \bigr) \ + \ {1 \over {2 r}} \ e^{-2 \phi} \
H^6 \wedge *H^6 \nonumber \\ 
&+& H \wedge {\tilde \omega^6} \ - \ H^6 \wedge {\tilde \omega}
\ + \ {1 \over 2} \ \omega \wedge \tilde{\omega}^6 \ - \ {1 \over 2} \ \omega^6 \wedge
\tilde{\omega} \quad .
\label{lagrred}
\ea
Here we use
slightly different conventions with respect to the previous section, 
namely a form language
and a space-time metric of signature $(5-1)$. $H^6$ and $\omega^6$ denote the internal
parts of the corresponding forms. Adding a  Lagrange multiplier $H^0 \wedge F = H \wedge F +
\omega \wedge F$  (locally, the solution for $F$ is $F=dC$), one gets 
\be
*H \ = \ * \bigl( H^6 \wedge Z \bigr) \ + \ {e^{2 \phi} \over r} \ {\hat F} \quad ,
\ee
where ${\hat F} = F - {\tilde \omega^6}$. The dualized Lagrangian is then
\ba
{\tilde {\cal L}}_{5} &=&
{1 \over 2r} \ e^{2\phi} \ {\hat F} \wedge *{\hat F} \ + \ {1 \over 2r} \ e^{-2\phi} \ 
H^6 \wedge *H^6  \ - \ H^6 \wedge Z \wedge {\hat F}  \nonumber \\
&-& \omega \wedge {\hat F} \ - \ H^{6} \wedge {\tilde \omega} \ - \ {1 \over 2} \
\omega \wedge  {\tilde \omega}^6 - {1 \over 2} \ \omega^6 \wedge {\tilde \omega} \quad .
\label{dualfive}
\ea

In order to compare ${\tilde {\cal L}}_{5}$ to the standard form of 
five-dimensional simple supergravity, one should make the following 
redefinitions of the  Abelian gauge fields: 
\ba
{\hat A}^{\rm x}_{\mu} &=& A^{\rm x}_{\mu} \ - \ a^{\rm 
x} \ Z_{\mu} \quad , \nonumber \\
{\hat  B}_{\mu} &=& B_{\mu 6} \ - \ v_{\rm xy} \ a^{\rm x} \ {\hat A}^{\rm 
y}_{\mu}   \quad , \nonumber \\
{\hat C}_{\mu} &=& C_{\mu} \ - \ {\tilde v}_{\rm xy} \ a^{\rm x} \ {\hat A}^{\rm 
y}_{\mu}  \quad .
\label{redef}
\ea
The resulting Lagrangian contains three types of Chern-Simons couplings,
\be
Z \ d {\hat B} \ d {\hat C} \qquad \quad {\hat C} \ v_{\rm x y} \ d {\hat 
A}^{\rm x} \ d {\hat A}^{\rm y} \quad \quad 
{\rm and } \quad \quad {\hat B} \ {\tilde v}_{\rm x y} \ d {\hat A}^{\rm x} 
\ d {\hat A}^{\rm y} \quad .
\label{cscoupl}
\ee 
They may be compared with the geometrical interaction 
(\ref{topcop}), and thus with the intersection numbers
$C_{\Lambda\Sigma\Delta}$ of the Calabi-Yau manifold. The $(n_T + n_V +1)$ 
scalar fields 
parametrize the hypersurface ${\cal V} = 1$, where ${\cal V}=C_{\Lambda\Sigma\Delta}
X^{\Lambda}  X^{\Sigma} X^{\Delta}$, with  $X^{\Lambda}(\phi)$  a set of $( n_T + n_V + 2 )$
special coordinates. In the case of a single tensor multiplet, 
the intersection form is
\be
{\cal V} \ = \ z \ b \ c \ - \ {1 \over 2} \ 
b \ {\tilde v}_{\rm xy} \ a^{\rm x} \ a^{\rm y}
\ - \ {1 \over 2} \ c \ v_{\rm xy} \ a^{\rm x} \ a^{\rm y} \quad .
\label{inters}
\ee
The generalization to $n_T > 1$ is fully determined by the $O(1,n_T)$ symmetry. Reverting
to the vector $O(1,n_T)$ notation of Section 2 and defining $b^r
=  ({{b + c} \over 2}, {{c - b} \over 2}, b^{r^{\prime}})$, with $( r^{\prime}=1,...,n_T - 1 )$,
one finally obtains
\be
{\cal V} = z \ b^r \ \eta_{rs} \ b^s \ - \ b^r \ c_{r {\rm xy}} \ a^{\rm 
x} \ a^{\rm y} \quad ,
\label{intscomp}
\ee
in terms of the special coordinates $X^{\Lambda} = (z, b^r, a^{\rm x})$. 
This structure of the intersection form is consistent with the condition 
that the manifold admits elliptic fibrations \cite{vafaf, morvafa}.

Cubic terms in any of the moduli should generally be allowed in eq.
(\ref{intscomp}), since they are compatible with five-dimensional supersymmetry. 
In six dimensions, these could correspond to a topological term  $\int F
\wedge F \wedge F$ for three six-dimensional vectors (see \cite{wittfm} for a recent discussion),
and would not contribute to  the field equations. The only prerequisite for having $a^3$ 
terms in $\cal V$ is thus the existence of a symmetric tensor $d_{\rm xyz}$ for 
the gauge group. In this case, the intersection form (\ref{intscomp}) can be
augmented by a term  $d_{\rm xyz} \ a^{\rm x} a^{\rm y} a^{\rm z}.$
On the other hand, the absence in six dimensions of cubic interactions for antisymmetric
tensors alone implies that in general eq. (\ref{intscomp}) cannot be 
modified by cubic terms in
the $b^r$ moduli, consistently with its general $O(1,n_T )$ structure.

Instanton contributions arising in the $S_1$ reduction from six 
to five dimensions may lead to additional terms in eq.
(\ref{intscomp}).  If these contributions are compatible 
with supersymmetry and modify the scalar kinetic terms,  
they must also modify the Chern-Simons couplings of
fields other than $b^r$ moduli. In the twelve-dimensional setting, 
these contributions can be seen to arise
from two-brane instantons \cite{bbs} when the two-brane is wrapped 
around 
$S_1 \times \gamma_2^{\alpha}$, where $\gamma_2^{\alpha}$ are two-cycles
on the CY manifold, excluding those that are on the base.

For $n_V =0$ the moduli space reduces to
\be
O(1,1) \times {{O(1,n_T )} \over {O( n_T )}} \quad ,
\ee
the original six-dimensional moduli space augmented with the radial mode.  In a similar
fashion, for $n_T=1$ and $\tilde{c}_{r \rm x y} = 0$, the space reduces to
\be
O(1,1) \times {{O(1,n_V + 1 )} \over {O( n_V + 1 )}} \quad ,
\ee
where the second factor is the Narain lattice \cite{narain} 
associated with the $S_1$ reduction of the dual heterotic theory on
$K_3$, while the
first factor is the six-dimensional moduli space for $n_T=1$.  When both 
$n_T \ge 1$ and $n_V > 0$, 
the space is no longer homogeneous and kinetic terms may have singularities corresponding to
phase transitions, thus extending the phenomenon of \cite{as,dmw}. This will
be discussed in more detail in the following section.

The final point we would like to stress is that, as discussed in the 
previous section,
the Lagrangian in eq. (\ref{dualfive}) is not invariant under the vector 
gauge
transformation. Consequently, the bosonic action obtained by reduction 
from six dimensions inherits an anomalous gauge variation.
In five dimensions, however, this anomaly can be canceled by a local 
counterterm $\int A^{\mu {\rm x}} J_{\mu {\rm x}}$, since the current
\be
* J_{\rm x} \ \sim \ c_{r {\rm xy}} \ c_{r {\rm zw}} \left( a^{\rm y} \  
F^{\rm z} \wedge F^{\rm  w} \ + \ 2 \ a^{\rm w} \ F^{\rm z} \wedge
F^{\rm y} \right) 
\label{current}
\ee
is gauge invariant.
\vskip 24pt
\section{Five-Dimensional Central Charges and Phase Transitions}

Phase transitions in six dimensions have received considerable 
attention lately, and have been studied extensively in 
\cite{sw,morvafa,dp,wittfm,aspinw,afiq,canf,klmvw}.  The analysis rests on 
the six-dimensional
formula for the central charge, uniquely determined by the $O(1,n_T;Z)$ duality symmetry,
\be
Z (\phi) \ = \ v^r ( \phi ) \ n_r \quad .
\label{central6}
\ee 
For $n_T = 1$, solving the constraint of eq. (\ref{scalars}) gives
\be
Z ( \phi ) \ = \ e^{\phi} \ n_e \ + \ e^{- \phi} \ n_m \quad ,
\label{z6one}
\ee
with a phase transition at $e^{-2 \phi} = - \ {{n_e} \over {n_m}}$.  Four-dimensional instanton
configurations of the gauge group $G_{\rm x}$ with instanton numbers $n_{\rm x}$ would give 
$n_e = n_{\rm x} \tilde{v}^{\rm x}$, $n_m = n_{\rm x} v^{\rm x}$, thus leading to a central charge
\be
Z ( \phi ) \ = \ n_{\rm x} \ ( \tilde{v}^{\rm x} \ e^{\phi} \ + \ v^{\rm x} \ e^{- \phi} ) \quad .
\label{centrinst6}
\ee
As discussed in detail in  \cite{sw}, $Z \rightarrow 0$ precisely when the gauge
kinetic term of eq. (\ref{vector}) becomes singular.  This is related to a new 
phenomenon, whereby a string becomes tensionless. 

In turning to five dimensions, six-dimensional strings may wrap around $S_1$, giving rise 
to point-like objects, or else they may simply reduce to five-dimensional strings.
The contribution of four-dimensional instantons to the five-dimensional charges may be
extracted from the equations
\ba
d * \left( {1 \over {\lambda}^2} \ e^{ - 2 \phi} \ d B \right) \ &=& \ 
{\tilde v}^{\rm x} \ Tr_{\rm x} ( F^2
)
\nonumber
\\ d * \left( {1 \over {\lambda}^2} \ e^{ 2 \phi} \ d C \right) \ &=& \ 
v^{\rm x} \ Tr_{\rm x} ( F^2 ) 
\quad ,
\label{fiveqs}
\ea
that follow from the dimensionally-reduced kinetic terms and 
from the Chern-Simons couplings,
once one sets $a^{\rm x} = 0$.  The relevant part of the action is then
\ba
\Delta {\cal L} &=& R \ + \ {1 \over 2} \ {\lambda}^4 \ dZ \wedge * dZ \ + \
{1 \over 2} \ \lambda \left( e^{- \phi} \ v^{\rm x} \ + \ e^{\phi} \ \tilde{v}^{\rm x} \right)
Tr_{\rm x} ( F
\wedge *F ) 
\nonumber \\
&+& { 1 \over {\lambda^2}}
 \left( e^{- 2 \phi} \ d B \wedge * d B \ + \ e^{2 \phi} \ dC \wedge * dC \right) \quad ,
\label{extract}
\ea
where $\lambda = r^{2/3}$ and $\phi$ denotes the six-dimensional dilaton.  The central charges for
electric (point-like) and magnetic (string-like) states  \cite{antft},
completely fixed by five-dimensional simple supergravity, are
\be
Z_e \ = \ X^{\Lambda} \ e_{\Lambda} \qquad \qquad Z_m \ = \ X_{\Lambda} \ m^{\Lambda}
\quad , \label{zem}
\ee
where $X^{\Lambda}$ are five-dimensional special coordinates and $X_{\Lambda}= C_{\Lambda
\Sigma \Delta} \ X^{\Sigma} X^{\Delta}$ are ``dual'' coordinates. 

Bearing in mind the structure of the intersection form (\ref{intscomp}),
at $a^{\rm x}=0$ the components of the gauge-field kinetic metric for $U(1)^3 \times \prod_{\rm x} G_{\rm
x}$  are
\be
G_{z z} = b^2 \ c^2 \qquad G_{b b} = {1 \over {b^2}} \qquad G_{c c} = {1 \over {c^2}}  
\qquad G_{{\rm x} {\rm x}} =c \ v^{\rm x} \ + \ b \ \tilde{v}^{\rm x} \quad .
\ee
Letting
\be 
b = \lambda \ e^{\phi} \qquad c = \lambda \ e^{- \phi} \qquad z={1 \over{b c}}={1 \over \lambda^2}
\quad ,
\label{defs}
\ee
one finds
\ba
Z_e &=& {1 \over \lambda^2} \ e_2 \ + \ \lambda \ \left( e^{\phi} \ e_1 \ + 
\ e^{-\phi} \ e_0 \right)
\ =
\ {1 \over{bc}} \ e_2 \ + \ b \ e_1 \ + \ c \ e_0 \nonumber \\
Z_m &=& \lambda^2 \ m_2 \ + \ {1 \over \lambda} \ \left( e^{-\phi} \ m_1 \ + \ 
e^{\phi} \ m_0 \right) \ =
\ b c \ m_2 \ + \ {1 \over b} \ m_1 \ + \ {1 \over c} \ m_0 \ . \label{zfivecen}
\ea

The non-perturbative phase-transition points are
 \ba
Z_e &=& 0 \qquad \left( e_2 = 0 , \quad \forall \lambda , \quad e^{2 \phi} = - {e_0 \over e_1} = - {v \over
\tilde{v}} \right) \nonumber \\
Z_m &=& 0 \qquad \left( m_2 = 0 , \quad \forall \lambda , \quad e^{2 \phi} = - {m_1 \over m_0} = - {v \over
\tilde{v}} \right) \quad . \label{cent5}
\ea

Thus, tensionless strings give rise to five-dimensional massless 
particles, as well as to tensionless
strings \cite{wittfm}. These are non-perturbative singularities occurring at 
non-vanishing instanton numbers.
In the zero instanton sector, $n_{\rm x} = 0$, one also expects perturbative 
BPS states that become massless:
\ba
Z_e &=& 0 \qquad \qquad \left( e_0 = 0 , \ \forall \phi , \ b^2 c = - 
{e_2 \over e_1} \right)
\nonumber
\\ Z_m &=& 0 \qquad \qquad \left( m_0 = 0 , \ \forall \phi , \  b^2 c
 = - {m_1 \over m_2}
\right) 
\quad .
\label{cent55y}
\ea

The non-perturbative singularities have an $F$-theory \cite{vafaf} 
interpretation in terms of the moduli space
of the base manifold, since they involve $\phi$, while the perturbative singularities
are associated with the fiber, since they involve the radial modulus 
$\lambda$. In the general case of
$a^{\rm x} \ne 0$ ($e^{\rm x}, m^{\rm x} \ne 0)$, one expects more complicated
singularities, which may also be  understood from the study of 
the CY threefolds.  Still, it is worth stressing 
that the general formula for the central charges, fully determined by
the intersection form (\ref{intscomp}), allows one to study these phase transitions
in arbitrary models in a rather general fashion.

\vskip 24pt
\section{Examples of Manifolds with $\chi = 0$}

As seen from eq. (\ref{euler}), a vanishing Euler characteristic implies that  
the number of tensor multiplets in six dimensions is $n_T=9$, while 
anomaly cancellation requires $n_H - n_V = 12.$ An interesting feature of 
these models is that the reducible part of the gravitational anomaly 
vanishes as well. Indeed, $I(R)=
I_{3 \over2} - (n_H + n_T - n_V)I_{1 \over2} - (n_T - 1)I_A$ vanishes 
identically 
when the coefficients of the second and third terms are 21 and 8, 
respectively \cite{alvwitt}.
We will consider  two such models that provide an interesting laboratory for 
testing duality conjectures.

Another motivation for considering these examples is to gain a better understanding of the 
connection between the present construction and $F$ theory that has already surfaced in
our discussion. Since our field content is the same as that
obtained from a CY compactification of $F$ theory, one
expects that the conditions for the existence of $F$ theory on the same 
CY manifold be encoded in the intersection form (\ref{intscomp}) that 
can be lifted to six dimensions.
It was shown in \cite{vafaf,morvafa} that, in order to obtain 
this six-dimensional spectrum, one must consider
compactifications of $F$ theory on manifolds that admit an elliptic fibration.
The structure of the intersection form obtained from $M$ theory
agrees with that result. Moreover, it displays the relation 
between the number $n_T$ of tensor multiplets and the number $k$ of K\"ahler 
deformations that does not change the K\"ahler class of the elliptic 
fiber \cite{vafaf} : $k = n_T +1$.  In eq. (\ref{intscomp}), the $b$ 
coordinates 
are moduli of the base manifold, $k=h^{1,1}(B)$, including the volume modulus 
that is in the hypermultiplet sector \cite{morvafa}. 
Note that the Hodge numbers of the CY manifold related to the 
five-dimensional model
have a six-dimensional interpretation, thus making it possible to lift 
to a hypothetical $F$
theory \cite {vafaf} in twelve dimensions.  Indeed, 
\be
h^{1,1} + h^{2,1} = n_T + 1 + n_V + n_H \quad ,
\label{count}
\ee 
where $( n_T +1 )$ is the number of (anti)self-dual tensors in six dimensions and $n_V$ and
$n_H$ are the numbers of vector multiplets and hypermultiplets, 
respectively. Moreover, \be
h^{2,1} = n_H -1
\label{counth}
\ee
since, as discussed in \cite{cadcdf}, one of the $h^{1,1}$ moduli, the 
CY volume deformation, is part of a universal hypermultiplet. 
In five dimensions the
universal hypermultiplet is really a linear multiplet, 
and comprises a three-form.
The arguments in the next section confirm this pattern in
six dimensions as well, where a four-form replaces one of the
scalars. This allows a natural generalization of 
the mechanism of ref. \cite{dsw} 
for the removal of one anomalous $U(1)$ 
via a coupling $A_4 \wedge F$.
\footnote{Actually, open descendants of Type-$IIb$ $K_3$
compactifications may accommodate a (model-dependent) number of four-forms 
originating from the ten-dimensional eight-form gauge potential (dual to
the $RR$ scalar) \cite{prep}. These can dispose of several anomalous 
$U(1)$'s, by means of couplings as above.}

\vskip.2in
\begin{flushleft}
{ \it The $(11, 11)$ model}
\end{flushleft}

This model was discussed in \cite{mirror} in the context of heterotic-Type $II$ 
duality, and more recently in \cite{morvafa}, in a related context.
A six-dimensional model with $n_T=9$ and $n_H =12$, a 
field content apparently in proper correspondence with the Hodge 
numbers, can be obtained from the Type-$IIb$ string compactified on 
$K_3$ with a freely acting $Z_2$. Presumably, it can also be  obtained 
from the completely higgsed $ (3, 243)$
model with one tensor multiplet un-higgsing eight 
tensors. Our
analysis is similar to the one in \cite{mirror}, where a CY manifold with
Hodge numbers $(11,11)$ was constructed as an orbifold of $K_3 \times T^2$.
Upon compactification on $K_3$, the Type $IIb$ string gives  an $N=(2,0)$ 
chiral theory in six dimensions with a supergravity 
multiplet that includes self-dual tensor fields in the 5 of $USp(4)$
coupled to 21 $N=2$ tensor multiplets \cite{seiberg}. 
In this model the scalar fields parametrize 
${{O(5,21)} \over {(O(5) \times O(21))}}$ \cite{rom}.  The theory can be truncated to 
$N=(1,0)$ resorting to the involution $\sigma$ of $K_3$. 
Under $Z_2$, $USp(4) \rightarrow USp(2) \times USp(2)$, with $4 \rightarrow
{(2,1)}^{+} + {(1,2)}^{-}$, and one of the gravitini
changes sign. Moreover, $5 \rightarrow (1,1)^+ + (2,2)^-$ , and since
$\sigma$  has eigenvalues $[(+1)^{10}; (-1)^{10}]$ when acting on $H^{1,1}$
and $-1$ when acting on $H^{(2,0)}$ and $H^{(0,2)}$ \cite{mirror}, $21 \rightarrow
9^+ + 12^-$. The moduli 
space of the $N=1$ theory is the subspace even under the involution: 
\be
{\cal M}_T \times {\cal M}_H =  {O(1,9) \over 
O(9)} \times {O(4,12) \over {O(4) \times O(12)} } \quad .
\label{moddd}
\ee
The result of \cite{mirror} 
that the five-dimensional moduli space for the vector 
multiplets (the K\"ahler moduli space of the Enriques surface at
fixed volume, together with the radial modulus) is
${\cal M}_V = O(1,1) \times {O(1,9) \over O(9)}$
can now be understood in terms of the $n_T =9$ tensor moduli space.
Note that the tensor multiplets in this 
model come from the untwisted sector. In contrast to the construction 
of \cite{cargese,bs,gp,dabpar}, here the $Z_2$ involution  
does not require the open-string sector, since in this case
$n_V=0$. The theory can be thought of 
as a compactification of $F$ theory on a CY manifold whose base is the Enriques 
surface ($h^{1,1}=10$) \cite{morvafa}, and in some sense is 
the simplest compactification of $F$ theory.  
 
Upon further reduction to five dimensions, one obtains a theory dual to $M$ theory on a
CY manifold with an intersection form 
\be
{\cal V} = z\ ( b \ c \ - \ b_{r^{\prime}} \ b^{r^{\prime}} )  \qquad (r^{\prime}=1,...,n_T -1)
\quad .
\label{intformbb}
\ee
The dual heterotic theory 
with $n_T =1$ and $n_V = 8$ \cite{mirror} is described by the same (exact) 
moduli space with an intersection form 
\be
{\cal V} = c(b \ z - a^{\rm x} \ a^{\rm y} \ d_{\rm xy} ) \quad .
\label{intformtt}
\ee
Note that in the latter case the $a^3$ term is absent, and we find 
a complete symmetry between these intersection forms. This symmetry interchanging vectors
from six-dimensional vector multiplets and vectors from six-dimensional tensor
multiplets is not present  in a generic five-dimensional theory.

\vskip.2in
\begin{flushleft}
{ \it The $(19, 19)$ model}
\end{flushleft}

Let us now turn to another model discussed in \cite{morvafa} - 
a manifold with $h^{1,1} = h^{2,1}=19$. The field content $n_T=9$, 
$n_V=8$ and $n_H=20$ has been obtained via an orbifold construction 
of $M$ theory\footnote{Other $n_T =9$ models may be obtained from 
$M$ theory on orientifolds of $K_3
\times S_1$ \cite{morb} and yield $n_V =  1,  2, 4$ and, correspondingly, $n_H = 13, 14, 18$. }
 \cite{sen} and as an orientifold of the Type-$IIb$ 
theory \cite{dabpar}. The model has a dual 
perturbative heterotic description \cite{papa} if the heterotic string is first 
compactified on $S_1$, so that the gauge group is broken to $U(1)^{16}$,
and is then compactified on $K_3$ to yield 17 vector multiplets and 20 hypermultiplets. 

The choice of a dilaton is naturally accompanied by 
the replacement of a pair of constrained tensors, its antiself-dual partner and
the self-dual one in the gravity multiplet, with a single unconstrained
tensor. After making a choice for the dilaton, one is thus left
with equal numbers of (anti)self-dual tensors and vectors. Still,  
the nature of the six-dimensional geometrical (and topological) interactions does
not allow a
symmetry in the intersection form between the $a$ and $b$ moduli. Thus, 
although in five dimensions tensor multiplets are identical
to vector multiplets, the theory ``remembers'' the origin of the vector fields.
It would be of interest to find models that are dual under the interchange
$(n_T -1, n_V) \leftrightarrow (n^{\prime}_T -1=n_V, n^{\prime}_V = n_T -1).$

\vskip 24pt
\section{Twelve-Dimensional Interpretation of Low-Energy Couplings}

In Section 2 we have discussed the six-dimensional origin of certain geometrical 
couplings:  upon reduction to five dimensions as in Section 3, they may be 
related to the intersection form of suitable (elliptically fibered) CY
reductions of $M$ theory.  The intersection forms are then in direct correspondence with
the $F_4 \wedge F_4  \wedge A_3$ geometrical interaction of eleven-dimensional supergravity
\cite{cjs}, the low-energy effective field theory of $M$ theory.  

In this section we would like to
return to our six-dimensional viewpoint in order to shed some 
light on the low-energy field equations of $F$ theory.
The resulting speculations are in the spirit of 
ref. \cite{cdfv}, where the 
Seiberg-Witten construction \cite{swym} 
was reinterpreted in terms of
a six-dimensional theory of (anti)self-dual tensor multiplets on 
${\cal M}_4 \times C_r$, with $C_r$ a genus-$r$ hyperelliptic Riemann 
surface. A recent realization of that proposal was presented in \cite{klmvw}.

Let us begin by considering the twelve-dimensional geometrical coupling
\be
T_{12} \ = \ \int_{{\cal M}_{12}} \ A_4 \wedge F_4 \wedge F_4 \quad ,
\label{t12}
\ee
where $A_4$ is the four-form potential of Type-$IIb$ ten-dimensional supergravity lifted
to twelve dimensions and $F_4$ is the four-form field strength of eleven-dimensional
supergravity lifted to twelve dimensions.  If we assume harmonic
expansions of $A_4$ and $A_3$ on $M_6 \times CY$ such that
\be
A_4 \ \supset  \ \sum_{r=1}^{h^{1,1}(B)=n_T+1} \ B_2^r \wedge {\tilde{V}_2}^r 
\quad , \label{a412}
\ee
with ${\tilde{V}_2}^r$ elements of $H^{1,1}(B)$, the cohomology
of the base, and
\be
A_3 \  \supset \ \sum_{{\rm x} = 1}^{h^{1,1} - h^{1,1}(B)-1=n_V} \ A_1^{\rm x} 
\wedge {\hat{V}_2}^{\rm x} \quad ,
\label{a312}
\ee
with the ${\hat{V}_2}^{\rm x}$ another subset of elements of the 
$H^2$ cohomology of the CY manifold,
eq. (\ref{t12}) induces six-dimensional couplings of the form
\be
T_6 \ = \ \int_{{\Sigma}_{6}} \ C_{r {\rm x y}} \ B_2^r \ \wedge F^{\rm 
x} \wedge F^{\rm y} \quad . \label{t6}
\ee
We thus learn that the coupling of eq. (\ref{t12}) must be present in $F$ theory.
Interestingly, this term joins the three-form of
eleven-dimensional supergravity and the four-form of ten-dimensional Type-$IIb$
supergravity, both lifted to twelve dimensions.  This also suggests that 
the five-brane of $M$ theory arises as a magnetic three-brane in $F$ 
theory, a property implicit in the coupling of eq. (\ref{t12}).  

The Chern-Simons coupling (\ref{t12}) induces currents for the 
twelve-dimensional gauge fields $A_3$ and $A_4$ of the form
\ba
&*&\!\!\! J_3 = F_4 \wedge F_5 \nonumber \\       
&*&\!\!\! J_4 = {1 \over 2} F_4 \wedge F_4 \quad .
\label{currents}
\ea
In the presence of five and six-brane sources,
these currents are not conserved,
since $d F_5 = Q^{(5)} \ \delta_6$ and $d F_4 = Q^{(6)} \ \delta_5$. 
However, current conservation is restored 
by additional contributions confined on the brane volumes, yielding
\ba
&*&\!\!\! J_3 = F_4 \wedge F_5 - Q^{(5)} \ A_3 \ \delta_6 + Q^{(6)} \ A_4 \ 
\delta_5 \nonumber \\    
&*&\!\!\! J_4 = {1 \over 2} F_4 \wedge F_4 - Q^{(6)} \ A_3 \ \delta_5 \quad .
\label{currents2}
\ea
The five-brane contribution comes from the world-volume coupling familiar 
from $M$ theory \cite{townsmm,wittenmm}. The six-brane coupling is expected
to come from the six-brane action coupled to the $A_3$, $A_4$ background
gauge fields.

The other terms needed to reproduce the six-dimensional couplings are
\be
L_1 \ = \ \int_{{\cal M}_{12}} \ A_4 \wedge I_8 ( R ) \quad ,
\label{l1}
\ee
and the topological terms
\ba
L_2 &=&  \int_{{\cal M}_{12}} \ F_4 \wedge I_8 ( R ) \quad , \nonumber \\
L_3 &=& \int_{{\cal M}_{12}} \ F_4 \wedge F_4 \wedge F_4 \quad .
\label{l3}
\ea
$L_1$ may be responsible for the six-dimensional tensor-gravity coupling, while the
topological terms $L_2$ and $L_3$ become gravitational and Chern-Simons \cite{dlm}  couplings of 
$M$ theory, once one takes ${\cal M}_{11} = \partial {\cal M}_{12}$. In five dimensions, $L_2$
describes the coupling of a linear combination of $n_V +1$ vectors to $TrR^2$ \cite{ferkm}, while
the reduction of $L_3$
to six dimensions gives rise to a term $d_{{\rm x y z}} \ a^{\rm x} \ a^{\rm y} \ a^{\rm z}$ in
the intersection form $\cal V$, as discussed in Section 3.

While all these terms may be regarded as a convenient bookkeeping for six-dimensional
couplings, their natural twelve-dimensional form is strongly suggestive of an $F$-theory
rationale for the CY compactification. In particular, the coupling of
eq. (\ref{t12}) implies that $F$ theory should accommodate both the 
Type-$IIb$ three-brane and the $M$-theory five-brane.  According to eqs. 
(\ref{a412})
and (\ref{a312}), however, $A_3$ and $A_4$ are not independent, since 
they are both
needed to get the entire $H^{1,1}$ cohomology.  Indeed\footnote{The
restrictions on the cohomological expansion of $A_3$ and $A_4$ may be 
regarded as 
induced from the yet unknown $F$-theory dynamics.}, only $H^{1,1}(B)$ 
contributes to $A_4$, while the rest contributes to $A_3$. Moreover,
$A_4$ should become self-dual when restricted to the Type-$IIb$ theory in ten dimensions.

Further confirmation of our proposal comes from $F$-theory 
compactified to eight dimensions on an elliptically fibered $K_3$ 
manifold. This yields the ``minimal'' eight-dimensional supergravity 
\cite{awtw}, dual to the one obtained from 
$M$-theory on $T_3/Z_2$, or equivalently from the heterotic string on $T_2$. 
In the supergravity 
multiplet the former contains a four-form gauge potential $A_{\mu 
\nu \rho \sigma}$, while the latter \cite{sasz} 
contains a dual two-form gauge potential $B_{\mu \nu}$.  This is possible thanks to the self-duality of 
the four-form field in ten dimensions, since $h^{1,1}(B)$ is one in 
this case. The scalar $\sigma$-model is $O(1,1) \times {O(2, n_V) \over {O(2) 
\times O(n_V)}}.$ The vectors of the eight-dimensional supergravity 
multiplet transform together with the $A^{\Lambda}_{\mu}$ of the 
vector multiplets under the duality group 
$G=O(2, n_V, Z)$, (with $n_V=18$) \cite{morvafa}, and indeed the theory has 
the geometrical interaction \cite{awtw}
\be
T_8 = C_{\Lambda \Sigma} \int_{{\cal M}_8} A_4 \wedge F^{\Lambda}_2 
\wedge F^{\Sigma}_2 \quad .
\label{eightint}
\ee
The twelve-dimensional coupling of eq. (\ref{t12}) 
reproduces precisely this term, if one assumes a harmonic expansion in the 
basis of the second cohomology of $K_3$ similar to (\ref{a312}):
\be
A_3 \ = \ \sum_{{\Lambda} = 1}^{b_2 - h^{1,1}(B)-1=n_V +2} \ A_1^{\Lambda} 
\wedge {\hat{V}_2}^{\Lambda} \quad .
\label{a3eq}
\ee
The intersection form is then $C_{\Lambda 
\Sigma} = [\Gamma_8 \oplus \Gamma_8 \oplus \sigma_1 \oplus \sigma_1]_{\Lambda 
\Sigma}$. According to our rule, the four-form gauge potential
is the only eight-dimensional remnant of $A_4$. In a similar fashion,
eq. (\ref{l1}) gives rise 
to a gravitational Chern-Simons coupling $\int_{{\cal M}_8} \ F_5 \wedge 
\omega_3 ( R )$ with a coefficient proportional to the Pontryagin number 
of $K_3$.

The previous arguments are in the spirit of the original 
motivation for $F$-theory as the underlying theory for the Type $IIb$ string.  
Since $B_{\mu \nu}$ and $B^c_{\mu \nu}$ form an SL(2,Z) doublet, 
related to the homology cycles of the two-torus, they are one-forms on 
$T_2$ and thus originate from a three-form in twelve dimensions. 
Taking into account that 
now $b_2 - h^{1,1}(B) -1 =0,$ there is only $A^+_4$ in ten dimensions, and 
the expansion of $A_3$ yields two antisymmetric tensors. The two scalars 
$(\phi, \phi^c)$ are related to the complex structure of torus. 
Once more, the Chern-Simons coupling (\ref{t12}) reproduces the Type $IIb$ 
couplings between the two- and four-form gauge potentials \cite{jhs},
aside from the self-duality of the five-form field strength. 

Finally, turning to the compactification of $F$-theory on a Calabi-Yau four-fold,  
recently considered in \cite{wittensp}, it is worth mentioning that in 
four complex dimensions the four-form field gives rise to a universal scalar 
degree of freedom $A_{ijkl} = C \epsilon_{ijkl}$ which may induce an 
(instanton-generated) superpotential by wrapping a three-brane 
world-volume on a four-cycle \cite{wittensp}. In addition, one could 
envisage another source for a potential - through the Chern-Simons 
term (\ref{t12}) if a four-form field strength condensate 
$F_{{\bar i}{\bar 
j}{\bar k}{\bar l}} \sim \epsilon_{{\bar i}{\bar j}{\bar 
k}{\bar l}}$, $F_{\mu \nu \rho \sigma} \sim \epsilon_{\mu \nu 
\rho \sigma}$ is allowed.

Let us conclude by observing that the twelve-dimensional couplings we have thus far
identified do not require a twelve-dimensional metric, a
pleasing feature since the twelve-dimensional theory cannot be an
ordinary gravitational theory.  Still, it is remarkable that some    
insight can be gained on the explicit couplings of $F$ theory from
these simple considerations.

\vskip 24pt
\begin{flushleft}
{\large \bf Acknowledgments}
\end{flushleft}

We would like to thank P. Berglund, M. Bianchi, A. Klemm, F. Riccioni 
and Ya.S. Stanev for interesting 
discussions. A.S. would like to thank the Theory Division of CERN for the
kind hospitality while this work was being completed.  
The work of S.F. was supported in part by
DOE grant DE-FG03-91ER40662, Task C., by EEC Science Program 
SC1$^*$CT92-0789 and by INFN. The work of R.M. is supported by a World  
Laboratory Fellowship.
The work of A.S. was supported in part by EEC Grant CHRX-CT93-0340.



\vskip 40pt

\begin{thebibliography}{99}
\bibitem{rom}{L.J. Romans, {\sl Nucl. Phys.} {\bf B276} (1986) 71.}
\bibitem{as}{A.~Sagnotti, {\sl Phys. Lett.} {\bf B294} (1992) 196.}
\bibitem{cargese} {A. Sagnotti, {\it in} ``Non-Perturbative Quantum 
Field Theory'',\\
eds. G. Mack et al (Pergamon Press, Oxford, 1988), p. 521.}
\bibitem{bs}{M. Bianchi and A. Sagnotti, {\sl Phys. Lett.} {\bf B247} (1990) 517;\\
{\sl Nucl. Phys.} {\bf B361} (1991) 519.}
\bibitem{dmw}{M.J.~Duff, R.~Minasian and E.~Witten, hep-th/9601036.}
\bibitem{sw}{N.~Seiberg and E.~Witten, hep-th/9603003.}
\bibitem{sen}{A.~Sen, hep-th/9602010.}
\bibitem{morb}{A. Kumar and K. Ray, hep-th/9602144.}
\bibitem{vafaf}{C.~Vafa, hep-th/9602022.}
\bibitem{morvafa}{D.R.~Morrison and C.~Vafa, hep-th/9602114, hep-th/9603161.}
\bibitem{cjs}{E. Cremmer, B. Julia and J. Scherk, {\sl Phys. Lett.} {\bf B76} (1978) 409.}
\bibitem{gst}{M. G\"unaydin, G. Sierra and P.K. Townsend, {\sl Nucl. Phys.} {\bf B242}
(1984) 244.}
\bibitem{wz}{J. Wess and B. Zumino, {\sl Phys. Lett.} {\bf B37} (1971) 95.}
\bibitem{wzsusy}{
G. Girardi, R. Grimm and R. Stora, {\sl Phys. Lett.} {\bf B156} (1985) 203;\\
L. Bonora, P. Pasti and M. Tonin, {\sl Phys. Lett.} {\bf B156} (1985) 341;\\
E. Guadagnini, K. Konishi and M. Mintchev, {\sl Phys. Lett.} {\bf B157} (1985) 37;\\
N.K. Nielsen, {\sl Nucl. Phys.} {\bf B244} (1984) 499;\\
H. Itoyama, V.P. Nair and H. Ren, {\sl Nucl. Phys.} {\bf B262} (1985) 317;\\
E. Guadagnini and M. Mintchev, {\sl Nucl. Phys.} {\bf B269} (1986) 543;\\
C.L. Bilchak, R. Gastmans and A. van Proyen, {\sl Nucl. Phys.} {\bf B273}
(1986) 46;\\
S. Ferrara, A. Masiero, M. Porrati and R. Stora,
{\sl Nucl. Phys.} {\bf B417} (1994) 238.}
\bibitem{gs}{M.B. Green and J.H. Schwarz, {\sl Phys. Lett.} {\bf B149} (1984) 117.}
\bibitem{type2b}{J.H. Schwarz, {\sl Nucl. Phys.} {\bf B226} (1983) 289; \\
P.S. Howe and P.C. West, {\sl Nucl. Phys.} {\bf B238} (1984) 181.}
\bibitem{ns}{H. Nishino and E. Sezgin, {\sl Nucl. Phys.} {\bf B278} (1986) 353.}
\bibitem{pss}{G. Pradisi, A. Sagnotti and Ya.S. Stanev, hep-th/9603097.}
\bibitem{tendsugra}{E. Bergshoeff, M. de Roo, B. de Wit and P. van
Nieuwenhuizen,\\
{\sl Nucl. Phys.} {\bf B195} (1982) 97;\\
G.F. Chapline and N.S. Manton, {\sl Phys. Lett.} {\bf 120B} (1983) 105.}
\bibitem{dp}{M.J. Duff, H. L\"u and C. Pope, hep-th/9603037.}
\bibitem{bz}{W.A. Bardeen and B. Zumino, {\sl Nucl. Phys.} {\bf B244} (1984) 421.}
\bibitem{dm}{M.J. Duff and R. Minasian, {\sl Nucl. Phys.} {\bf B436} (1995) 507.}
\bibitem{cadcdf}{A.~C.~Cadavid, A.~Ceresole, R.~D'Auria and S.~Ferrara,
{\sl Phys. Lett.} {\bf B357} (1995) 76.}
\bibitem{antft}{I.~Antoniadis, S.~Ferrara and T.R.~Taylor, {\sl Nucl. Phys.}
{\bf B460} (1996) 489.}
\bibitem{ferkm}{S.~Ferrara, R.R.~Khuri and R.~Minasian, hep-th/9602102.}
\bibitem{wittfm}{E.~Witten, hep-th/9603150.}
\bibitem{bbs}{K.~Becker, M.~Becker and A.~Strominger, {\sl Nucl. Phys.}
{\bf B456} (1995) 130.}
\bibitem{aspinw}{P.S. Aspinwall, hep-th/9511171.}
\bibitem{afiq}{ G. Aldazabal, A. Font, L.E. Ibanez
and F. Quevedo, hep-th/9602097.} 
\bibitem{canf}{P. Candelas and A. Font, 
hep-th/9603170.} \bibitem{klmvw}{A. Klemm, W. Lerche, P. Mayr, C. Vafa and
N. Warner, hep-th/9604034.}
\bibitem{alvwitt}{L.~Alvarez-Gaum\'e and E.~Witten, {\sl Nucl. Phys.} {\bf 
B234} (1983) 269.}
\bibitem{dsw}{E. Witten, {\sl Phys. Lett.} {\bf B149} (1984) 351;\\
M. Dine, N. Seiberg and E. Witten, {\sl Nucl. Phys.} {\bf 
B289} (1987) 589.} 
\bibitem{prep}{C. Angelantonj, M. Bianchi, G. Pradisi, 
A. Sagnotti and Ya.S. Stanev, in preparation.}
\bibitem{mirror}{S.~Ferrara, J.A.~Harvey, A.~Strominger and C.~Vafa,
{\sl Phys. Lett.} {\bf B361} (1995) 59.}
\bibitem{papa}{G.~Papadopoulos and P.K.~Townsend, {\sl Phys. Lett.} {\bf B357} (1995) 
300.}
\bibitem{seiberg}{N. Seiberg, {\sl Nucl. Phys.} {\bf B303} (1988) 286.}
\bibitem{gp}{ E.C. Gimon and J. Polchinski, hep-th/9601038.}
\bibitem{dabpar}{A.~Dabholkar and J.~Park, hep-th/9602030.}
\bibitem{narain}{K.S. Narain, {\sl Phys. Lett.} {\bf B169} (1986) 41.}
\bibitem{cdfv}{A. Ceresole, R. D'Auria, S. Ferrara and A. Van
Proeyen,\\ {\sl Nucl. Phys.} {\bf B444} (1995) 92.}
\bibitem{swym}{N. Seiberg and E. Witten, {\sl Nucl. Phys.} {\bf B426} 
(1994) 19 \\ {\sl Nucl. Phys.} {\bf B431} (1994) 484.}  
\bibitem{townsmm}{P.K.~Townsend, hep-th/9512062.} 
\bibitem{wittenmm}{E.~Witten, hep-th/9512219.}
\bibitem{dlm}{M.J.~Duff, J.T.~Liu and R.~Minasian, {\sl Nucl. Phys.} 
{\bf B452} (1995) 261.}
\bibitem{awtw}{M. Awada and P.K. Townsend, {\sl Phys. Lett.} {\bf B156} 
(1985) 51.} 
\bibitem{sasz}{A. Salam and E. Sezgin, {\sl Phys. Lett.} {\bf B154} 
(1985) 37.}
\bibitem{jhs}{J.H. Schwarz, {\sl Nucl. Phys.} {\bf B226} (1983) 269.}
\bibitem{wittensp}{E.~Witten, hep-th/9604030.}

\end{thebibliography}
\end{document}